\documentclass[pre,notitlepage,superscriptaddress,showpacs,showkeys,longbibliography,twocolumn]{revtex4-1}
\usepackage{latexsym,amssymb,amsfonts,mathrsfs,amsmath,bm}
\usepackage{graphicx}
\usepackage[usenames,dvipsnames]{xcolor}
\usepackage{soul}
\usepackage{float}
\usepackage{bbold}
\usepackage[linktocpage,colorlinks=true,linkcolor=blue,citecolor=blue,breaklinks=true,urlcolor=blue]{hyperref}
\usepackage{textcomp}

\newcommand{\bff}{{\bf f}}
\newcommand{\bu}{{\bf u}}

\newcommand{\bx}{{\bf x}}

\newcommand{\uvc}[1]{\bm{\mathrm{\hat #1}}} 


\begin{document}

\title{A multiscale biophysical model gives quantized metachronal waves in a lattice of cilia}  
\author{Brato Chakrabarti}
\affiliation{Center for Computational Biology, Flatiron Institute, New York, NY 10010, USA}
\author{Sebastian F\"{u}rthauer}
\email{sfuerthauer@flatironinstitute.org}
\affiliation{Center for Computational Biology, Flatiron Institute, New York, NY 10010, USA}
\author{Michael J. Shelley}
\email{mshelley@flatironinstitute.org}
\affiliation{Center for Computational Biology, Flatiron Institute, New York, NY 10010, USA}
\affiliation{Courant Institute, New York University, New York, NY 10012, USA}
\date{\today}

\begin{abstract}
\noindent Motile cilia are slender, hair-like cellular appendages that spontaneously oscillate under the action of internal molecular motors and are typically found in dense arrays. These active filaments coordinate their beating to generate metachronal waves that drive long-range fluid transport and locomotion. Until now, our understanding of their collective behavior largely comes from the study of minimal models that coarse-grain the relevant biophysics and the hydrodynamics of slender structures. Here we build on a detailed biophysical model to elucidate the emergence of metachronal waves on millimeter scales from nanometer scale motor activity inside individual cilia. Our study of a 1D lattice of cilia in the presence of hydrodynamic and steric interactions reveals how metachronal waves are formed and maintained. We find that in homogeneous beds of cilia these interactions lead to multiple attracting states, all of which are characterized by an integer charge that is conserved. This even allows us to design initial conditions that lead to predictable emergent states. Finally, and very importantly, we show that in nonuniform ciliary tissues, boundaries and inhomogeneities provide a robust route to metachronal waves. 
\end{abstract}

\maketitle


\noindent Motile cilia are thin hair-like cellular
projections that serve as fundamental building blocks of locomotion and material transport in many eukaryotes. Each cilium is an active machine driven by thousands of internal nanometric molecular motors that collectively conspire to produce whip-like oscillations along its $\sim10\mu\mathrm{m}$ length. Cilia often coat tissues in dense carpets, with these carpets producing collective metachronal waves (MWs) that travel across these ciliated surfaces of length scales over $\sim 100 \mu\mathrm{m}$ or more. How these waves result from the regulation and coordination of biophysical dynamics over length scales spanning six orders of magnitude is the question that we seek to address in this paper.


Metachronal waves are ubiquitous in nature. In mammals, ciliated tissues circulate cerebrospinal fluid in the brain \cite{faubel2016cilia,pellicciotta2020entrainment}, pump
mucous, and remove foreign particles trapped in lung airways
\cite{sleigh1988propulsion}. In humans, their dysfunction underlies
diverse pathologies \cite{reiter2017genes}. Metachronal waves are responsible
for the locomotion of ciliated unicellular organisms such as
\textit{Paramecium} \cite{tamm1972ciliary} and \textit{Volvox}
\cite{brumley2012hydrodynamic} and serve as feeding and filtering
pumps for marine invertebrates \cite{gilpin2017vortex}. Additionally, in many vertebrates, motile cilia are involved in symmetry breaking during embryonic development \cite{shinohara2012two,smith2012symmetry,r2019cilium}. 

While the appearance of MWs across various systems is
extremely robust, the microscopic physics and interactions that result
in their emergence are not fully understood \cite{byron2021metachronal}. Plausible physical
mechanisms include mechanical coupling through the anchoring membrane
\cite{machemer1972ciliary}, local steric interactions within dense arrays
\cite{chelakkot2021synchronized}, and large-scale fluid motion that
induces long-range coupling among cilia \cite{gueron1997cilia}. 
Many models coarse-grain the
internal mechanics and the fluid-structure interactions by approximating cilia
as spheres driven on compliant orbits
\cite{lagomarsino2003metachronal,niedermayer2008synchronization,goldstein2009noise,uchida2010synchronization,golestanian2011hydrodynamic,wollin2011metachronal,brumley2012hydrodynamic,brumley2015metachronal}. Hydrodynamic interactions are sufficient to explain the collective behavior in such minimal models for
ciliary arrays \cite{uchida2010synchronization,nasouri2016hydrodynamic,mannan2020minimal,liao2021energetics,hamilton2021changes}.
However, it remains unclear to which extent these conclusions are robust in the context of the intricate biophysics and hydrodynamics that govern actual ciliated surfaces. The few direct simulations that resolve the beating dynamics of an individual cilium
\cite{gueron1997cilia,gueron1998computation,osterman2011finding,elgeti2013emergence,ding2014mixing}
either neglect or coarse-grain \cite{guirao2007spontaneous,yang2008integrative,han2018spontaneous,martin2019emergence} the
internal biomechanics of spontaneous oscillations. The present study aims at bridging this gap. 

Here, for 1D lattices of cilia, we identify four crucial ingredients that lead to the robust emergence of MWs. First is the spontaneous oscillation of a single cilium; for this 
we build upon a biophysical model of a single cilium 
\cite{oriola2017nonlinear,chakrabarti2019spontaneous,chakrabarti2019hydrodynamic},
that incorporates many essential features of the microscopic actuation physics. 
Second and third are hydrodynamic and steric interactions which lead to the coordination between cilia. Fourth are spatial inhomogeneities, or lack thereof, that dictate the allowable states of organization in a ciliary bed. 

\begin{figure*}[t]
	\centering 
	\includegraphics[width=1\linewidth]{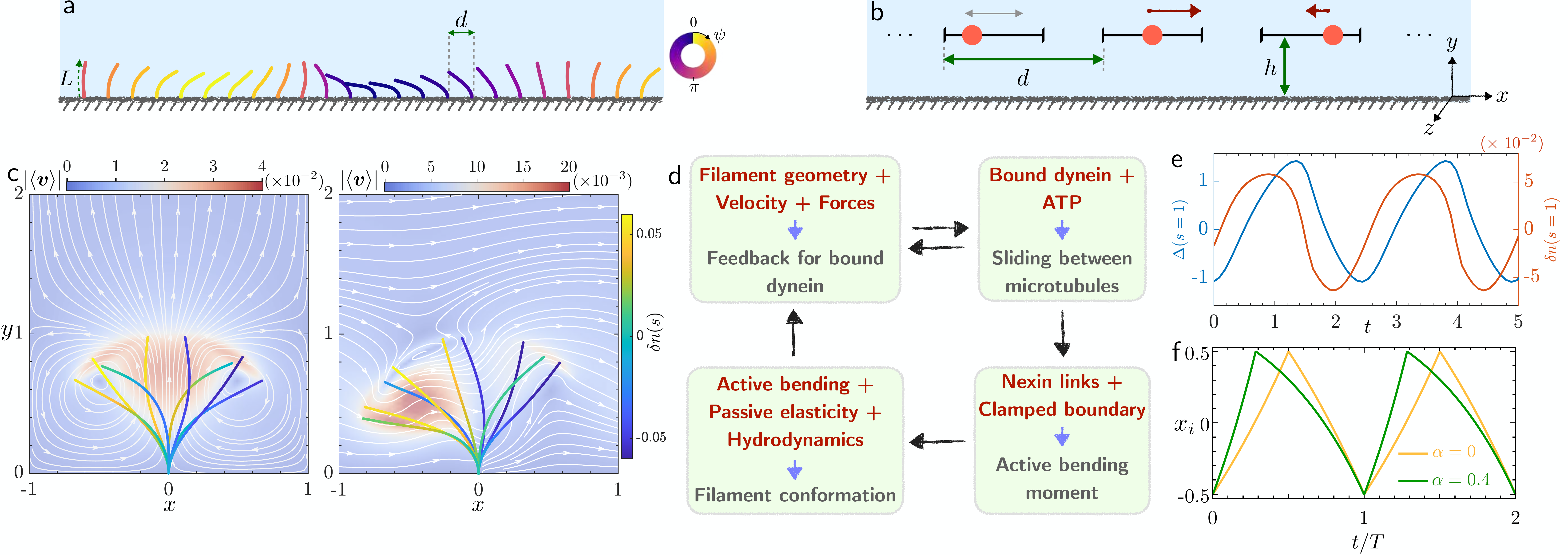}\vspace{-0.3cm}
	\caption{\textbf{\textsf{Individual beating patterns, flow-fields, and MWs on a lattice of active filaments and rowers.}}  
    {\textsf{\textbf a,}} Snapshot of an emergent metachronal wave from simulations of active filaments on a 1D lattice. Identical filaments of length $L$ are arranged with a fixed lattice spacing $d$ above a no-slip wall. The filaments beat inside a 3D bath of an incompressible Newotnian fluid that allows the flow to go around them. The filaments interact via long-range hydrodynamic interactions and pairwise steric repulsion. The properties of the MWs are governed by the ratio $d/L$, fluid viscosity, and the internal mechanics of spontaneous oscillations. Within one complete time-period, we can assign a unique phase $\psi(t) \in (0, 2\pi]$ to each filament conformation. The filaments on this figure are colored according to their instantaneous phase $\psi(t)$.
	{\textsf{\textbf b,}} Arrangements of \textit{rowers} on a 1D lattice that serve as a coarse-grained description of the active filaments. Each filament is represented by a spherical bead (rower) moving on a 1D track at a height $h$ above a no-slip wall. The tracks are separated by a distance $d$. Every rower is driven by a constant force. The arrows on the spherical beads qualitatively represent the magnitude and direction of this force.
	{\textsf{\textbf c,}} We show the beating patterns of individual active filaments for two different sets of internal parameters. We overlay various filament conformations over one period. The filaments are color-coded by the instantaneous motor distribution $\delta n(s,t) = n_+-n_-$ along their centerlines. The activation and inhibition of the two groups of motors ($n_{\pm}$) during the beating can be observed.
	On left, we have fore-aft symmetric and on right we have asymmetric beating patterns with distinct power and recovery strokes. In the background, we display the period-averaged velocity fields and streamlines for each of these cases. The flow field inside the box is computed with periodic boundary conditions along the $x$ direction. For asymmetric beating (right) it highlights a net fluid pumping in the $x$ direction (see movies Mov\_01 and Mov\_02 in the SM).
	{\textsf{\textbf d}}, Schematic of the geometric-feedback loop that governs the dynamics of the active filaments. This loop results in a Hopf bifurcation leading to spontaneous oscillations.
	{\textsf{\textbf{e}}}, The evolution of the sliding displacement $\Delta(s,t)$ at the tip $s=1$, between microtubule doublets caused by the dynein motors is shown from the steady state of asymmetric beating. The evolution of $\delta n(s,t)$ indicates that there is a constant time-delay between the action of motor proteins and the response of a cilium.
   {\textsf{\textbf f,}} Oscillations of a single rower over two time periods $T$ are shown. Depending on the specific choice of the driving force we can either have left-right symmetric ($\alpha = 0$) or asymmetric ($\alpha \neq 0$) oscillations.}
   \label{fig:lattice}
\end{figure*}


We show that in homogeneous ciliary beds, hydrodynamic and steric interactions give rise  to multiple attracting states of which metachronal waves constitute a small portion. Guided by coarse-grained models we find that all such states are characterized by an invariant integer charge. We further find that heterogeneities in the ciliary bed break charge conservation.  This leads to robust formation of metachronal waves, independent of initial conditions, in finite beds and patchy lattices of cilia. These results provide a rigorous demonstration of how spontaneous beating of cilia driven by internal molecular motors, hydrodynamic and steric interactions, and the morphology of ciliary beds, together shape metachronal waves.


\vspace*{1mm}
\noindent\textbf{\textsf{A hierarchy of models}}


\noindent Our work is centered around a detailed biophysical model of individual active filaments \cite{chakrabarti2019hydrodynamic}, many of which are coupled together hydrodynamically and sterically. To better identify and understand the role of various interactions that regulate the dynamics of ciliary arrays, we also use a hierarchy of more coarse-grained mathematical models, the first of which is the  \textit{rower model}. It shares similarities with many previously studied minimal descriptions of cilia and will help us elucidate the role of long-range hydrodynamic interactions. Next, we outline a \textit{phase dynamics model} and its continuum analog \cite{kuramoto2003chemical}. This helps explain the role of interactions that prevent sharp phase-gradients on a lattice. 



\noindent \textit{A detailed biophysical model:} The internal core of a cilium is the axoneme with a diameter $a \sim 200 \mathrm{nm}$. This
highly conserved structure consists of 9 pairs of microtubule
doublets arranged circularly about a pair of microtubules \cite{gilpin2020multiscale}.  Typical cilium lengths vary
from $L \sim 8-15 \ \mu m$ \cite{gueron1997cilia} and we describe
this slender structure by its centerline $\bx(s,t)$, parametrized by arclength $s$. In our active filament model \cite{chakrabarti2019spontaneous}, the axoneme is modeled
as two shearable, polar, planar elastic rods confined in the $x$-$y$ plane and clamped at their bases. These two rods, $\bx_{\pm} = \bx(s,t) \pm a \uvc{n}(s,t)/2$, represent the microtubule doublets from the opposite sides of the axoneme where $\uvc{n}(s,t)$ is the unit normal to the centerline $\bx(s,t)$. The dynein motors together with passive nexin cross-linkers between microtubule doublets generate shear forces $\bff^{\pm}_m(s)$ per unit length. This leads to a sliding displacement $\Delta(s,t) = a \int_0^s \|\bx_{ss}(s',t)\| \mathrm{d} s'$ between the two rods. The force density can then be expressed as
\begin{equation}
\bff^{\pm}_{m}(s, t)= \pm \bx_s \left[\rho\left(n_{+} F_{+}+n_{-} F_{-}\right)-K \Delta(s,t) \right].
\end{equation}
where $\rho$ is the line density of motors, $n_{\pm}$ is the fraction of bound motors on $\bx_{\pm}$, $F_{\pm}$ is the force exerted by an individual dynein, and $K$ is the stiffness of nexin links modeled as linear springs.
The force exerted by the motors follows a linear force-velocity relation $F_{\pm} = \pm f_0 (1 \mp \Delta(s,t)_t/v_0)$, where $f_0$ is  the stall force of dynein and $v_0$ is a characteristic velocity scale. The lack of sliding at the base means that the sliding forces are converted into an active bending moment
\begin{equation}
\mathbf{M}(s,t) = B \bx_s \times \bx_{ss} - \uvc{z} a \int_s^L \| \bff^{\pm}_m(s',t) \| \ \mathrm{d}s',
\end{equation}
where $B$ is the bending resistance. The evolution of the centerline $\bx(s,t)$ follows from  nonlocal slender-body-theory (SBT):
\begin{equation}
\partial_t \bx(s,t) = \mathcal{M}(\bx(s,t),d) \cdot \bff(\bx(s,t)),
\end{equation}
where $\mathcal{M}$ is an integral kernel accounting for anisotropic drag and hydrodynamic interactions with other cilia \cite{tornberg2004simulating}. The force density $\bff$ exerted by the filament on the fluid has contributions from active moments, tensile forces, and bending deformations. Along with the hydrodynamics we also account for pairwise repulsive forces that prevent overlap between neighboring filaments on a lattice (see Methods). 

A key feature of our model is the regulation of the force-generating dynein  population through feedback from the filament deformations determined by hydrodynamic stresses. This feedback will ultimately allow the filaments to spontaneously beat in isolation and to coordinate their beat patterns. The bound motor population follows a first-order kinetic equation $\partial_t n_{\pm} = \pi_{\pm} - \epsilon_{\pm}$ with
\begin{align}
\pi_{\pm} &= \pi_0 (1-n_{\pm}), \\
\epsilon_{\pm} &= \epsilon_0 n_{\pm} \mathcal{P}_{\pm} (F_{\pm}) ,
\end{align}
where $\pi_0,\epsilon_0$ are characteristic rate constants. The geometric feedback arises from a force-dependent detachment of motors modulated through the function $\mathcal{P}_{\pm}$ (see Methods and Fig.~\ref{fig:lattice}\textsf{\textbf d}). 

Depending on the specific choices for $\mathcal{P}_{\pm}$ an isolated cilium can have two qualitatively different beating patterns. First, the filaments can have left-right symmetric oscillations. This does not break time-reversal symmetry \cite{purcell1977life}, and so does not pump fluid on average (see Fig.~\ref{fig:lattice}\textsf{\textbf c}(left)). This  fundamentally differs from cilia, that typically beat asymmetrically with distinct power and recovery strokes. We build this biologically relevant behavior by an appropriate choice of $\mathcal{P}_{\pm}$ that amounts to introducing phenomenological asymmetries into the axoneme \cite{chakrabarti2019spontaneous,chakrabarti2019hydrodynamic}. 
With this, we obtain whip-like beating and a non-zero mean flow over one time period (Fig.~\ref{fig:lattice}\textsf{\textbf c}(right)). 
We also note from Fig.~\ref{fig:lattice}\textsf{\textbf f}, that only a small fraction of the bound motor population $\approx 5\%$ is sufficient to drive these oscillations \cite{oriola2017nonlinear}.
\begin{figure*}[t]
	\centering 
	\includegraphics[width=1\linewidth]{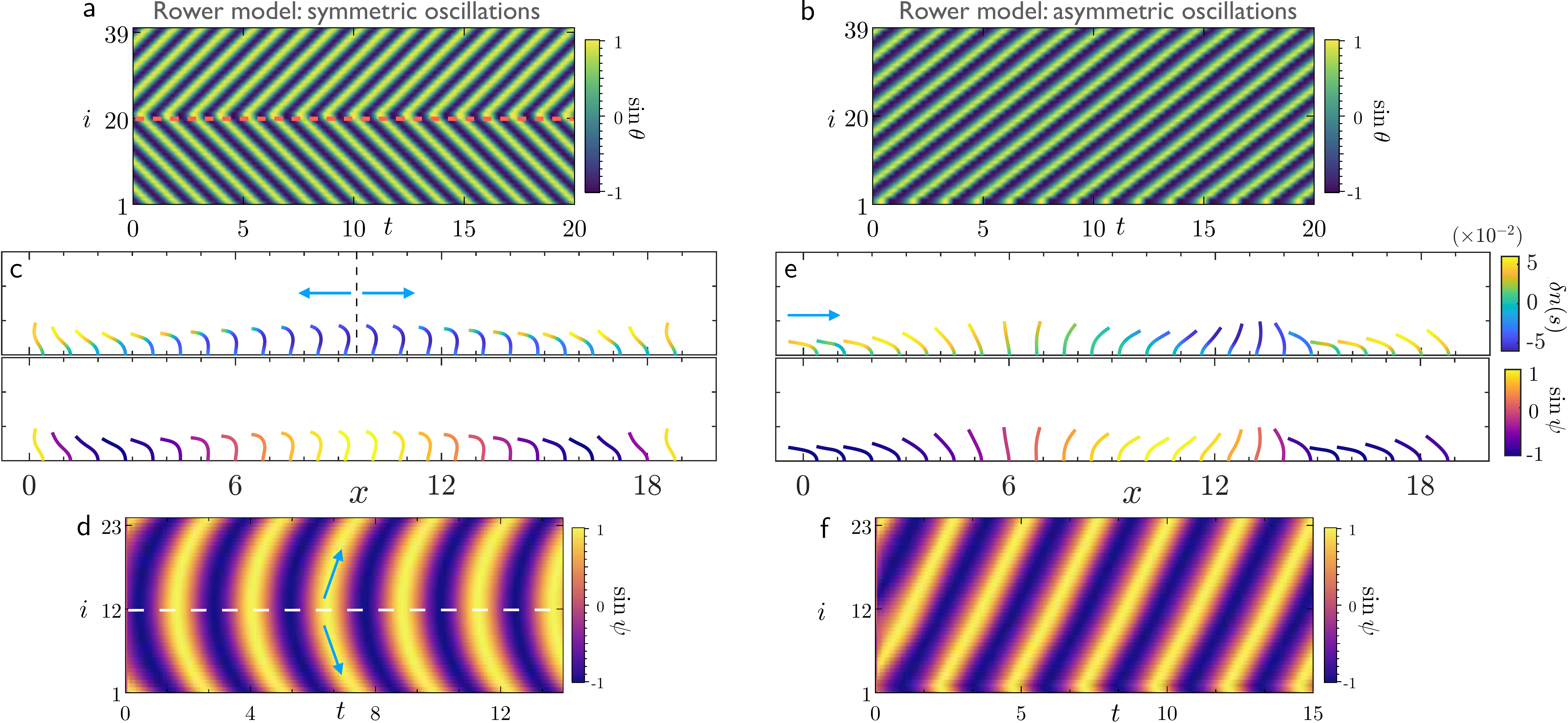}
	\vspace*{-0.5cm}
	\caption{\textbf{\textsf{Asymmetric beating pumps fluid in naturally emerging MWs on finite beds of filaments.}}  
	{\textsf{\textbf a,}} We consider a lattice of 40 identical rowers with open fluid on either side. Each rower is identified by their unique phase $\theta_i \in (0,2\pi]$ (see Methods). We display a kymograph of phases from the emergent state of rowers for symmetric ($\alpha = 0$) oscillations. Irrespective of the initial conditions, the final state is characterized by a wave that splits from the middle of the domain (red dashed line). The wave then propagates in both directions of the lattice with identical wave speeds (see movie Mov\_03 in the SM).
	{\textsf{\textbf b,}} For asymmetric ($\alpha = -0.3$)  rowers, the final state is characterized by a symplectic metachronal wave that spans the entire domain (see movie Mov\_05 in the SM). 
	\textsf{(\textbf {c,d}),} Symmetric beats of active filaments also result in wave splitting from the middle of the domain. \textsf{\textbf{c}} shows a snapshot of filament conformations from the final state of wave splitting. On the top row, each filament centerline $\bx(s,t)$ shows the distribution $\delta n(s,t) = n_+ - n_-$. This serves as a measure of activity, and highlights the coordinated action of dynein across the lattice. In the bottom row, the filaments are color-coded by their instantaneous phase $\psi_i$. On \textsf{\textbf d} we display the corresponding kymographs of phase. The arrows indicate the direction of wave propagation (see movie Mov\_04 in the SM). \textsf{\textbf e,} Identical to the rower model, asymmetric beating of active filaments results in a symplectic metachronal wave. This wave propagates along the lattice, in the direction indicated by the arrow resulting in pumping of fluid. The filament conformations are from the emergent steady state of wave propagation. The associated phase kymograph is shown in \textsf{\textbf f} (see movie Mov\_06 and Mov\_07 in the SM). To make distinction between the active filaments and rowers we will use different color codes to represent their respective phase kymographs. Parameters: For rowers, $h = 0.4$, $d = 1.3$, $k_e = -0.6$, $N=40$; active filaments, $d/L = 0.8$, $N=24$ in \textsf{\textbf{c}}. For \textsf{\textbf{e}}, we are displaying $N=24$ filaments from a simulation of $N=100$ filaments.} 
	\label{fig:openwave}
\end{figure*}

\vspace*{1mm}

\noindent \textit{Rowers:} Our first coarse-grained minimal description for hydrodynamically interacting active filaments is the well known \textit{rowers model}  \cite{wollin2011metachronal,guo2018bistability,hamilton2021changes} that exhibit a rich variety of dynamics. Here we represent each filament by a sphere of radius $a$, identified by position $x_i$ moving on a 1D track above a height $h$ from a no-slip wall (see Fig.~\ref{fig:lattice}\textsf{\textbf b}). Each sphere (or rower) also has a state $\sigma_i = \pm 1$ and is driven by a horizontal force $F_i = F_0 \sigma (1+ \alpha \sigma) - k_e x_i$, such that
\begin{equation}\label{eq:rowev}
\dot{x}_{i}= \xi(h) F_i + \sum_{j \atop j \neq i} \mathcal{G}(d,h) \cdot F_{j}, \ \ \ i = 1,2, \cdots N
\end{equation}
where $\xi(h)$ is the hydrodynamic mobility of the sphere, and $\mathcal{G}(d,h)$ is the hydrodynamic interaction kernel. Once the rower reaches the end of its track $\sigma$ abruptly changes its sign reversing the direction of the driving force and resulting in sustained oscillations. The parameter $\alpha$ gives us the provision to modulate the driving force $F_i$ such that it has  different magnitudes depending on the sign of $\sigma$. As shown in Fig.~\ref{fig:lattice}\textsf{\textbf e}, for $\alpha \neq 0$ oscillation of a rower is left-right asymmetric over a period. Finally, $k_e < 0$ is the curvature of a harmonic potential that has its maximum at the midpoint of the tracks. The specific choice of $k_e$ destabilizes the midpoint. Since all rowers are hydrodynamically coupled, their beat periods can vary allowing them to spontaneously synchronize. 


\vspace*{1mm}

\noindent \textit{Phase dynamics model:}  Metachronal waves are fundamentally related to coordinated phase dynamics of limit cycle oscillators for which the cilium is a prototypical example. An array of identical, interacting oscillators with phase $\psi_i$ and intrinsic frequency $\omega$ obeys 
\begin{equation}\label{eq:phase}
\dot{\psi_i} = \omega + \varepsilon \left[f \left(\psi_i-\psi_{i-1} \right) + f \left(\psi_i-\psi_{i+1} \right)\right].
\end{equation}
This description is generic in the limit of weak coupling ($\varepsilon \ll 1)$, where a formal averaging procedure \cite{kuramoto2003chemical,pikovsky2003synchronization}  allows one to coarse-grain any interactions in terms of coupling functions $f(x)$. Here $f(x)$ is $2 \pi$-periodic. 
While our model is a variant of the classical Kuramoto  oscillator \cite{kuramoto2003chemical}, where $f(\psi_i-\psi_j) = \sin (\psi_i-\psi_j)$, our subsequent discussions are  independent of the  specific choice of the coupling function.
\begin{figure*}[t]
	\centering 
	\includegraphics[width=0.8\linewidth]{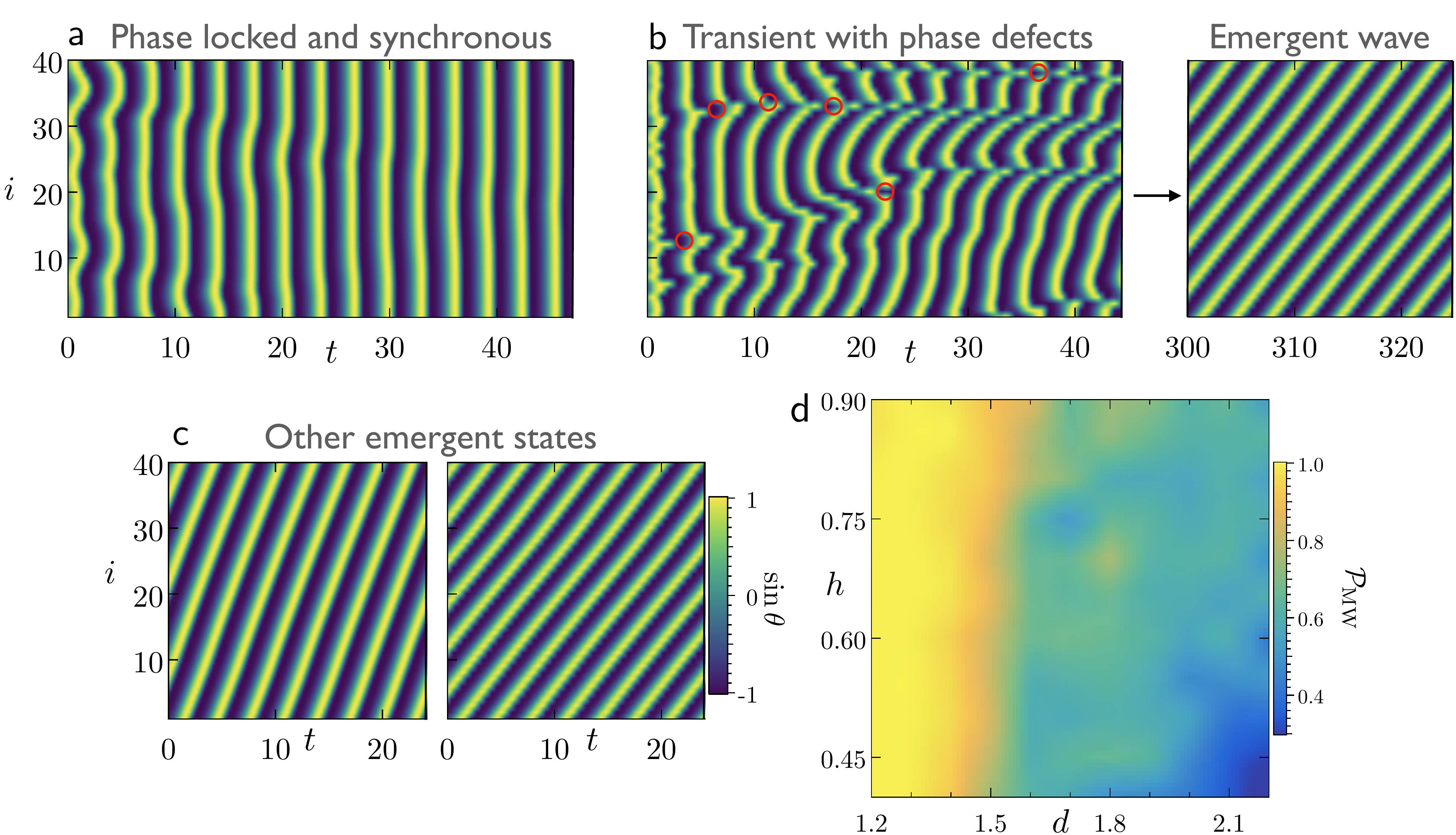}\vspace{-0.3cm}
	\caption{\textbf{\textsf{Multiple attracting states and their basin of attraction in a homogeneous bed of rowers.}}  In {\textsf{(\textbf{a-c}),}} we consider identical internal parameters for rowers interacting on a lattice with periodic boundary conditions. We study the evolution of their phase starting from different random initial conditions. {\textsf{\textbf a,}} The rowers evolve towards a phase locked state where all the oscillators are phase synchronized and have an identical frequency. \textsf{\textbf b,} Symplectic metachronal waves emerge on the lattice. Formation of the MW is characterized by transient phase defects at early time. Some of these defects are marked with red circles. \textsf{\textbf c,} Multiple MW solutions are possible for the final state. We show two such final states with different wave speeds and wavelengths. The properties of the MWs depend on the initial conditions of rowers (see movie Mov\_08 in the SM). \textsf{\textbf d,} We characterize the basin of attraction of MW solutions as a function of lattice spacing $d$ and  height $h$. To this end, we compute the probability of formation of MWs ($\mathcal{P}_{\text{MW}}$) by averaging over 50 initial conditions for each pair of $(d,h)$. In a dense lattice of rowers MWs are the dominant attractor. However, $\mathcal{P}_{\text{MW}}$ decreases with increasing lattice spacing. The basin of attraction is sensitive to the internal parameters of the model and MWs constitute a relatively small portion of the entire parameter space (see SM for another example). Parameters: In all simulations, $\alpha =-0.3$, $k_e = -0.6$, and $N = 40$. For \textsf{(\textbf{a-c})} we chose $h = 0.4$, $d = 1.3$.}
	\label{fig:rowper}
\end{figure*}

\vspace*{1mm}
\noindent\textbf{\textsf{Finite beds robustly generate MWs}}

\noindent We start by looking at emergent dynamics in finite beds of active filaments and rowers. In such a setup, the boundaries of the arrays are open. In the context of our phase dynamics model, this amounts to oscillators at the edge $(i=1, N)$ interacting with only one neighbor. Figure~\ref{fig:openwave} illustrates the phase dynamics of ciliary arrays with a small number of active filaments, and rowers.  Independent of initial conditions, we find that both the active filaments, and the rowers, adjust their phases and evolve to a unique state characterized by propagating waves.  

For symmetric oscillators, the final state is characterized by a wave that splits in the middle of the domain and propagates to both sides with identical wave speeds; see Fig.~\ref{fig:openwave}(\textsf{\textbf {a,c,d}}). However, on introducing asymmetry into the beating patterns, we find the emergence of unidirectional metachronal waves spanning the whole array, see Fig.~\ref{fig:openwave}(\textsf{\textbf{b,e,f}}). The latter is the case which we think is most relevant for biological cilia. Figure~\ref{fig:openwave}(\textsf{\textbf{c,e}}) illustrates the coordination of molecular motors in the lattice where we have color coded each filament centerline $\bx(s,t)$ by the instantaneous distribution $\delta n(s,t) = n_+-n_-$ which  serves as a measure of activity. Associated with each of these filament conformations is a unique phase $\psi(t)$ that can be used to characterize their synchronization (see Figure~\ref{fig:openwave}(\textsf{\textbf{c,e}})(bottom row)). Our results reveal that phase dynamics in a MW is indeed a result of spatio-temporal self-organization of motor proteins across the lattice. The waves in our problem are symplectic as they propagate in the direction of mean fluid transport (see SM). The characteristics of these waves are controlled by lattice spacing and the details of the internal mechanics of the symmetry-breaking oscillations. The rowers recapitulate all the behaviors of the complex biophysical model, highlighting that hydrodynamic interactions alone can sustain metachronal waves. 


\begin{figure*}[t]
	\centering 
	\includegraphics[width=1\linewidth]{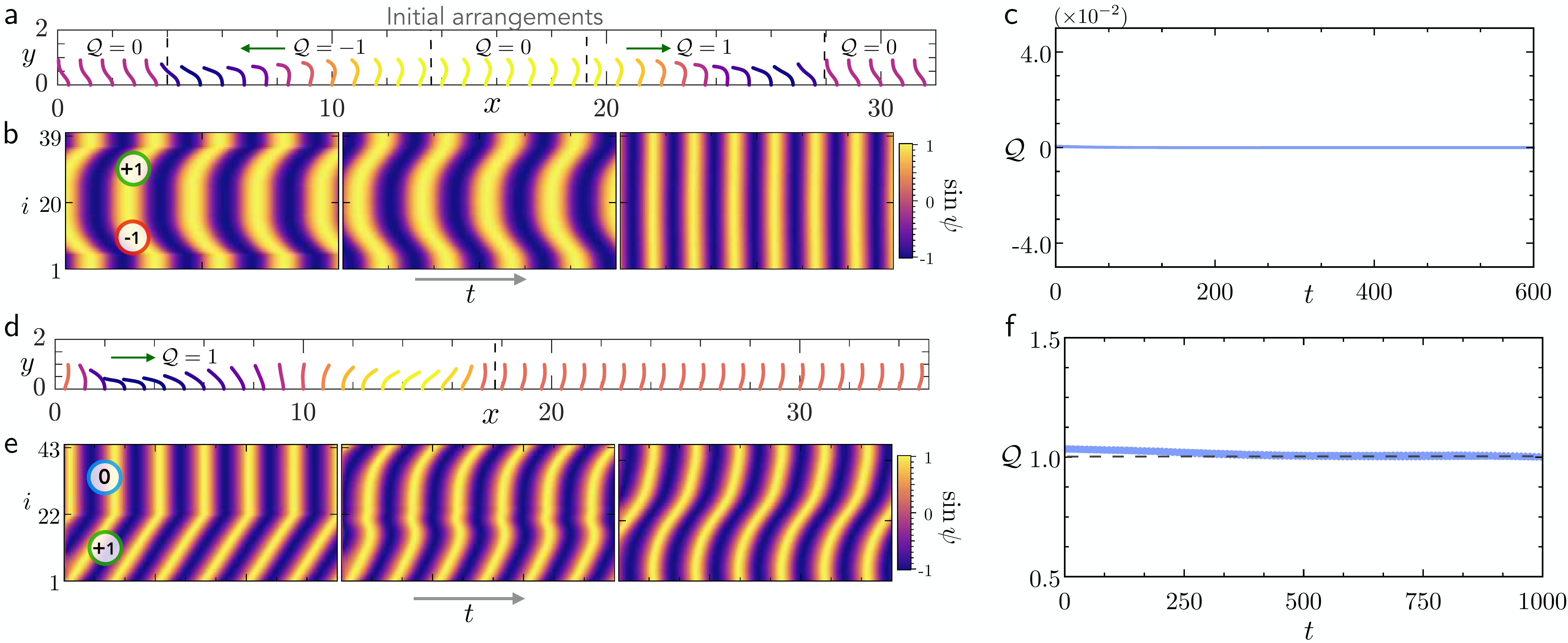}\vspace{-0.3cm}
	\caption{\textbf{\textsf{Attractors in homogeneous beds of active filaments are quantized by an integer charge.}} ({\textsf{\textbf{a,d}}}) We first illustrate our method of \textit{`cut-out-and-stitch-in'} to design initial conditions. The subplots indicate the initial filament conformations where every filament is colored by its instantaneous phase $\psi_i$. The vertical lines are the boundaries of different regions that are stitched together to form the initial state. The green arrows show the initial direction of wave propagation by the stitched regions. We also indicate the charge carried by the different parts of the initial state.
	({\textsf{\textbf{b,e}}}) We show the evolution of the kymographs of filament phases. On the initial kymograph we have shown the charge from different stitched regions. We notice that as time progresses the phase fronts are smoothed out. This is indicative of the diffusive dynamics of the phase.
    ({\textsf{\textbf{c,f}}}) Evolution of the charge $\mathcal{Q}$ over time. The charge is approximately conserved during the evolution and the emergent state is characterized by its integer value. Parameters: In all the examples, $d/L = 0.8$; \textsf{\textbf a} corresponds to symmetric and \textsf{\textbf{d}} to asymmetric oscillations.}
	\label{fig:deswave}
\end{figure*}

\vspace*{1mm}
\noindent\textbf{\textsf{MWs are non-generic in homogeneous beds}}

\noindent Ciliated tissues typically contain thousands of cilia at high density \cite{gilpin2017vortex}. A commonly used mathematical abstraction for such a large-scale system is a bed of filaments with periodic boundaries \cite{elgeti2013emergence}. 
We first study this scenario through our minimal model of rowers. The evolution towards MWs is always characterized by the appearance of phase defects due to locally anti-phase oscillations at early times, some of which are marked in red circles on Fig.~\ref{fig:rowper}\textsf{\textbf{b}}. Importantly, Fig.~\ref{fig:rowper}(\textsf{\textbf{a,b}}) highlights that randomly initialized rowers can lead to both phase-locked state and metachronal waves for an identical set of parameters \cite{hamilton2021changes}. Thus, unlike finite beds of cilia, metachronal waves are not generic in this configuration. 

To gain further insight consider the evolution of the phase dynamics model
\begin{align}\label{eq:phaseper}
&\dot{\psi}_{i} =\omega +\varepsilon f\left(\psi_{i}-\psi_{i+1}\right)+\varepsilon f\left(\psi_{i}-\psi_{i-1}\right), 
\end{align}
In a infinite or periodic homogeneous bed, all the oscillators are coupled identically. The model \eqref{eq:phaseper} have fixed points for which the phase difference $\delta_i = \psi_i-\psi_{i+1}$ are an arbitrary constant $\delta_i = \delta$. The case of $\delta=0$ corresponds to a phase-locked synchronous state with $\psi_i = \Psi(t)$. All the other cases result in formation of MWs with different wavelengths. Figure~\ref{fig:rowper}\textsf{\textbf c} shows two such potential MW states whose property is solely determined by the initial conditions. The qualitative features of the dynamics are unaltered for symmetric rowers ($\alpha = 0$) where waves can propagate in either direction.


Our numerical exploration suggests that many of these fixed points are attractors, a feature that has recently been reported in other minimal models of hydrodynamically coupled cilia \cite{meng2020designing,solovev2020global}. 
In Fig.~\ref{fig:rowper}\textsf{\textbf d} we characterize the basin of attraction of possible MWs as a function of the lattice spacing $d$ and height $h$ of the rowers. We find that for a dense bed of rowers, MWs span almost the entire space of initial conditions. However, for larger lattice spacing the system tends to evolve towards the synchronous fixed-point more often. The boundaries of the attractors are sensitive to the internal parameters (see SM).
 \begin{figure*}[t]
 	\centering 
 	\includegraphics[width=0.97\linewidth]{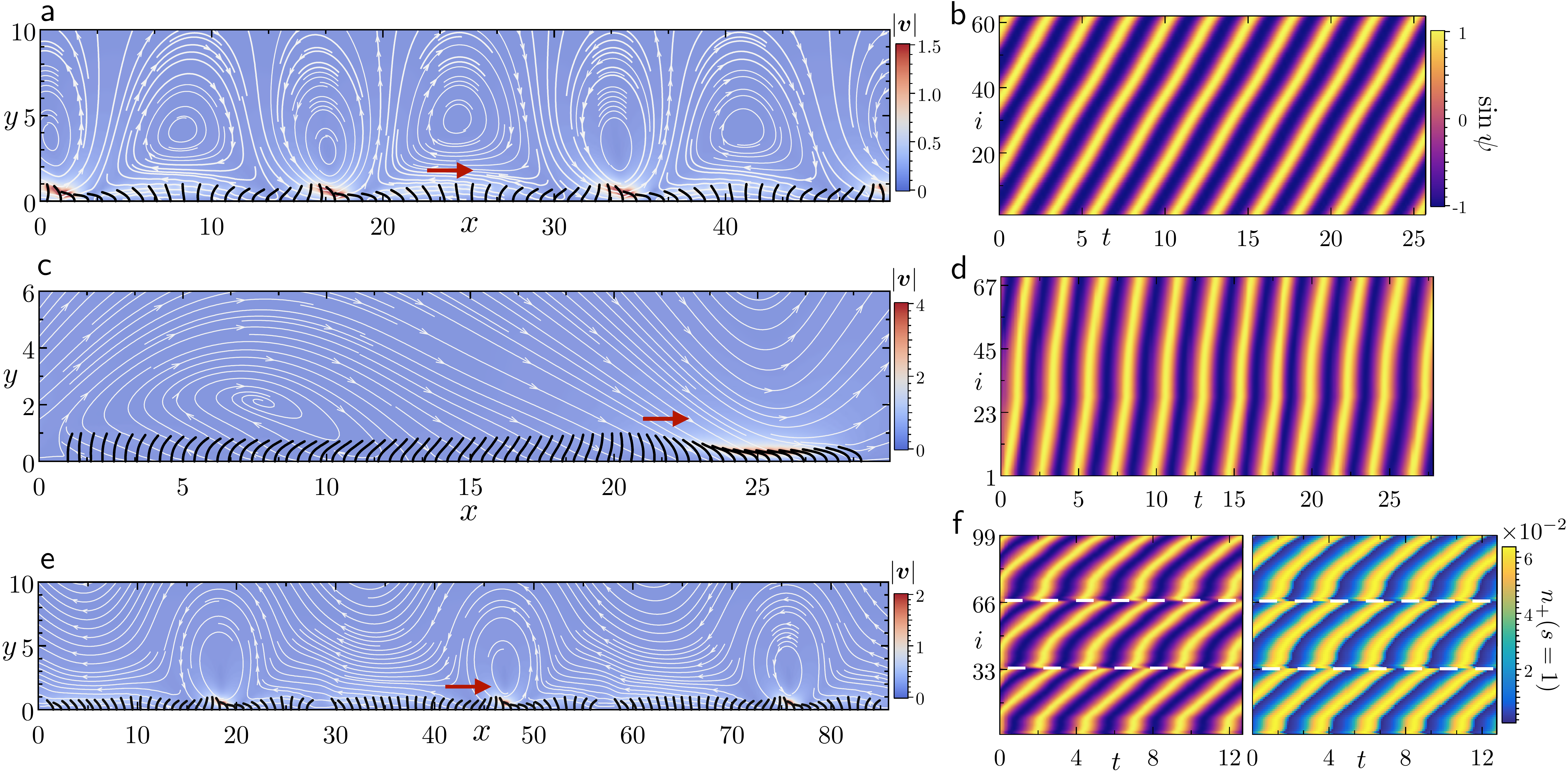}\vspace{-0.3cm}
 	\caption{\textbf{\textsf{Self-organization of dynein and robust MWs in heterogeneous ciliary beds.}}  {\textsf{\textbf a,}} We show instantaneous filament conformations from the final state of steady MWs in a periodic lattice with $N = 62$ active filaments. The arrow indicates the direction of wave propagation. This particular state is characterized by charge $\mathcal{Q} \approx 3$. In the background, we have displayed the instantaneous flow-field that contains large-scale vortical structures. The associated kymographs of the phases from the final state is shown on \textsf{\textbf{b}}  (see movie Mov\_09 in the SM). 
     (\textsf{\textbf{c-d}}) Periodic lattice with a non-uniform arrangement of $N=70$ active filaments. This leads to the robust formation of MWs. We display the filament conformations from the final state, and the associated kymographs of the phase. The arrow indicates the direction of wave propagation (see movie Mov\_10 in the SM). \textsf{\textbf e}, Emergent  MWs in patches of ciliated cells inside a periodic lattice. Each patch of 33 active filaments  generates its own MW. These waves subsequently coordinate over long time. The filament conformations are from the final steady-state of wave propagation. \textsf{\textbf f}, The left figure shows the associated kymographs of phase. The white lines indicate the boundaries of individual patches. On right, we show a kymograph of the fraction of bound motor population $n_+$ at the tip of the filaments. We note that the kymograph of $n_+$ has the same structure as the phase. This highlights the self-organization of nanometric machines to coordinate and drive metachronal waves across millimetric length scales (See movie Mov\_11 in the SM). Parameters: $d/L = 0.8$ for (\textsf{\textbf{a,e}}) and $d/L = 0.4$ for \textsf{\textbf c}.}
 	\label{fig:patch}
 \end{figure*}

\vspace*{1mm}
\noindent\textbf{\textsf{Coarse-grained PDE predicts quantized states}}

\noindent To better understand this let us consider the continuum limit of our phase dynamics model where we assume that phases $\psi_i$ vary smoothly along the lattice. When the spacing between the oscillators $\Delta x \to 0$ with $\varepsilon (\Delta x)^2 \equiv \tilde{\varepsilon}$  finite, it is possible to obtain a coarse-grained PDE by expanding the coupling function $f(x)$  \cite{pikovsky2003synchronization}. The phase evolution is governed by
\begin{equation}\label{eq:phapde}
\partial_t \psi = \omega + \gamma \partial^2_x \psi + \beta \left(\partial_x \psi \right)^2, 
\end{equation}
where $\gamma = \tilde{\varepsilon}f'(0)$, and $\beta = \tilde{\varepsilon}f''(0)$. It follows from Eq.~\eqref{eq:phapde} that the phase-gradient $\partial_x \psi$ obeys a diffusive Burgers' equation. This coarse-grained description is strictly valid for wavelengths larger than $\Delta x$.  

In a periodic lattice of length $\ell_B$, the phase satisfy $\psi(x) = \psi(x+\ell_B+2 \pi n)$, where $n \in \mathbb{Z}$. Under the evolution described by Eq.~\eqref{eq:phapde} we have a conserved quantity: namely, the charge $\mathcal{Q}$ defined as
\begin{equation}\label{eq:charge}
\mathcal{Q} = \frac{1}{2 \pi} \int_0^{\ell_B} \partial_x \psi \ \mathrm{d} x.
\end{equation}
The charge is an integer and characterizes the winding number of the phase \cite{shraiman1992spatiotemporal} in the box. Using the definition of the charge $\mathcal{Q}$ along with Eq.~\ref{eq:phapde} we find that $\mathrm{d}\mathcal{Q}/\mathrm{d}t = 0$, and thus the charge is conserved during the evolution of the phase. Importantly, for our continuum description, this means that an initial state with charge $\mathcal{Q}$ will generate a steady state solution that conserves this charge. We next ask, to what extent this insight translates to our rower model, and then to our biophysical model of active filaments.  


As shown in Fig.~\ref{fig:rowper}\textbf{\textsf{b}}, the transient dynamics of hydrodynamically interacting rowers can develop regions where nearby oscillators are anti-phase. This violates the assumptions of the continuum theory developed above and the charge $\mathcal{Q}$ is no longer conserved (see SM). In contrast, we find that pairwise steric repulsion and hydrodynamic interactions mediated by contact of filaments in our biophysical model prevent the appearance of sharp phase-gradients and suggest charge conservation. 





\vspace*{1mm}
\noindent\textbf{\textsf{Designing waves  in periodic filament lattices}}

\noindent We propose a method of \textit{`cut-out-and-stitch-in'} to test conservation of charge. If charge in the biophysical model is conserved, we should be able to design initial conditions that give rise to predictable emergent states. In Fig.~\ref{fig:deswave}\textsf{\textbf a}, we start with an array of 40 filaments with symmetric beating patterns. We leverage filament conformations from emergent waves on finite beds to design initial states with different charges. For symmetric beating, we \textit{cut-out}  two waves that propagate in opposite directions and  \textit{stitch-in} with them additional regions where filaments are synchronous. The kymograph associated with this initial state is shown in Fig.~\ref{fig:deswave}\textsf{\textbf b}, where we have indicated the charge carried by the different regions of the lattice. The net charge in the initial state is close to zero. As the system evolves, we note that the initially sharp phase fronts in the kymograph are smoothed out. This is a signature of diffusive dynamics. Finally, the system evolves into a phase-locked state with synchronous beating. The charge $\mathcal{Q}$ shown in Fig.~\ref{fig:deswave}\textsf{\textbf c} is seen to be approximately conserved and is close to zero as the system evolves and reaches the synchronous beating state. 
Figure~\ref{fig:deswave}\textsf{({\textbf{d-e}})} showcases another example for asymmetric beating of filaments relevant to ciliary  beds. In Fig.~\ref{fig:deswave}\textsf{\textbf d} we have stitched together a metachronal wave carrying unit charge with synchronous states to create a state with $\mathcal{Q} \approx 1$. The diffusive dynamics is evident in the smoothing of the phase kymographs and $\mathcal{Q}$  is approximately preserved during the evolution as seen in Fig.~~\ref{fig:deswave}\textsf{\textbf f}. These examples elucidate the existence of multiple attracting states in homogeneous beds of filaments and highlights that such states are quantized by the charge $\mathcal{Q}$. Figure~\ref{fig:patch}\textsf{\textbf a} shows a  traveling wave characterized by $\mathcal{Q}\approx3$. The filament conformations from the final state are overlaid on an instantaneous flow field that features large-scale vortical structures. We conclude from these numerical explorations that in the full system, charge is to a good approximation conserved. We attribute the slight differences to our definition of phase and the finite size of the system. Above, we speculate that charge conservation in the biophysical system is a consequence of short-range interactions that prevent sharp phase-gradients. This seems plausible since these interactions give rise to the same phenomena in a different internal mechanics model of active filaments (see SM). Thus it should be possible to break the conservation law by introducing controlled spatial inhomogeneities into the ciliary bed. We next test this assertion numerically.

Figure~\ref{fig:patch}\textsf{\textbf c} depicts a specific design of a periodic ciliary lattice, in which active filaments are distributed non-uniformly to introduce systematic heterogeneity, which is relevant to all biological systems \cite{pellicciotta2020entrainment}. Similar to finite beds of cilia, this configuration robustly gives rise to metachronal waves from arbitrary initial conditions. Importantly, this means that charge conservation can be broken in a controlled way by introducing spatial inhomogeneities. We next test this in a biologically plausible morphology.  

Ciliary carpets in airways of vertebrates are characterized by patchy distribution of cilia \cite{holley1986alignment,francis2009initiation}.
Recent experiments on mouse trachea have revealed that this heterogeneous arrangement of cilia is crucial for  long-range transport \cite{ramirez2020multi}. In the context of our problem, this patchiness provides a natural way to introduce heterogeneity which hints at additional role of non-uniformities. Like in the case shown in Fig.~\ref{fig:patch}\textbf{\textsf{c}}, a patchy filament distribution in a periodic lattice results in a generic attracting state of MWs. We show this in Fig.~\ref{fig:patch}(\textsf{\textbf{e-f}}), where we consider three patches of ciliated surface, each containing 33 filaments and placed inside a periodic lattice. For a cilia of $L \sim 12 \mu \text{m}$ and $d/L = 0.8$ this lattice spans  distances of $\mathcal{O}(\mathrm{mm})$. As highlighted by the kymograph from the final state in Fig.~\ref{fig:patch}\textsf{\textbf{f}}(left), each of these patches develops its own metachronal wave that eventually coordinates across the patches over long time scales. Finally, we want to emphasize that our simulations rigorously bridge length scales spanning six orders of magnitude in a biologically relevant morphology. In Fig.~\ref{fig:patch}\textsf{\textbf{f}}(right), we show the strong correlation between the coordination at the nanometric length scales that result in emergent dynamics spanning millimeters. 

\vspace*{1mm}
\noindent \textbf{\textsf{Discussion}}


\noindent We have used three mathematical models of varying complexity to elucidate the microscopic physics underlying the emergence of metachronal waves on a 1D lattice of cilia (see SM for tabulated summary). Our detailed biophysical model provides us a unique insight into the role of dynein motors in the collective dynamics. Figure~\ref{fig:patch}\textsf{\textbf f}(right) shows the fraction of bound motors $n_+$ at the tip of every filament as a function of time. Since the spontaneous oscillations of these filaments are driven by the action of dynein, they share the same structure as the kymographs of the phase. At the nanometric length scale, hydrodynamic interactions through the chemo-mechanical feedback loop  coordinate the operation of these motors, resulting in self-organization at millimetric length scales.

 
Both our rower and biophysical model point at existence of multiple steady state solutions in homogeneous beds, commonly used to model large systems. In our active filament simulations there are both steric and strong hydrodynamic interactions mediated through contact or near contact of the filaments in dense arrays. In contrast to the rowers, these interactions in our biophysical model prevent sharp phase-gradients on the lattice (see Mov\_12 in SM). This allows us to test the predictions of the continuum phase dynamics model which assumes smooth variation of the phase along the lattice and predicts that all allowable states should be characterized by an invariant integer charge $\mathcal{Q}$. We indeed find that the phase evolution of the biophysical model on periodic lattices yield states quantized by the charge $\mathcal{Q}$. This provides us with a method to control and design final attracting states. 

An important question in Biology is how ciliated tissues control their collective beat patterns. In \cite{solovev2020global} the authors argue that biochemical noise might determine the emergent states. Our work points at another possibility. We show that spatial inhomogeneities establish a robust pathway for waves to arise on patchy ciliated surfaces. This also has further important biological implications. Our findings imply that patched ciliated tissues are not only better at generating flow-fields that help in long-range transport \cite{ramirez2020multi}, but they also help individual patches to become better pumps by self-organizing into metachronal waves. 

The extension of our predictions to two-dimensional sheets of cilia is a natural and important question. The role of steric and hydrodynamic forces that regulate filament dynamics in close contact is sensitive to the ciliary beating plane. Their ability to ensure smooth variations of phase dynamics across a 2D lattice remains unclear. In fact, interactions can potentially lead to disorganized coordination as recently reported in orthoplectic rows of hydrodynamically coupled cilia with planar beating patterns \cite{han2018spontaneous}. Thus, the possibility of quantized states in a 2D lattice is likely to hinge upon internal mechanics of ciliary beating not yet understood. Moreover, realistic ciliated tissues pose several sources of heterogeneity, ranging from uneven cilia spacing and variation in beating planes \cite{ramirez2020multi} to deviations in intrinsic ciliary beat frequency \cite{pellicciotta2020entrainment}, roles of which remain to be understood.  

Our simulation framework incorporates the full microscopic details of beating cilia. Here it has allowed us to understand how ciliary bed morphology, beating patterns, steric, and hydrodynamic interactions work together to shape the emergent dynamics on 1D lattices. We believe that this is a foundational step towards understanding how the properties of motor proteins shape ciliary beats and collective dynamics in their natural settings.

\footnotesize
\section*{Methods}
\noindent \textsf{\textbf{Active filament model and simulations:}} Our model for the spontaneously beating cilium closely follows \cite{chakrabarti2019spontaneous,chakrabarti2019hydrodynamic,chakrabarti2019problems}. Here we provide a concise summary of the model and relevant equations for consistency. Each filament is indexed by $i$ and its centerline is identified by a Lagrangian marker $\bx_i(s,t)$ which is parameterized by the arc length $s$. The end $s=0$ is clamped against a no-slip wall and $s=L$ is free. The centerline dynamics of the $i^{\mathrm{th}}$ filament in the array follows from  slender-body theory (SBT) as
\begin{align}
\begin{split}
8 \pi \nu \left[\partial_t \bx_i(s,t) - \bu^d(\bx_i(s,t))\right] &=  \mathcal{M} \cdot \bff_i(s,t) \label{eq:SBT} 
\end{split}
\end{align}
where $\nu$ is the viscosity of the fluid and  $\bff_i$ is the force per unit length exerted by the filament on the fluid. $\mathcal{M}$ is the operator for  SBT \cite{tornberg2004simulating,chakrabarti2019hydrodynamic} which can be written as
\begin{equation}
\mathcal{M}[\bff] = \Lambda[\bff] + \mathcal{K}[\bff].
\end{equation}
The first term accounts for local anisotropic drag that depends on the aspect ratio of the filament \cite{chakrabarti2019spontaneous} and the second term accounts for non-local interactions \cite{tornberg2004simulating}. In \eqref{eq:SBT} $\bu^d_i$ is the disturbance velocity generated by all the other filaments at $\bx_i(s,t)$. This accounts for the long-range hydrodynamic interactions and is given by
\begin{equation}
\bu^d_i(\bx_i(s,t))= \sum_{\substack{j=1 }}^N \int_0^L \mathsf{G}^{\epsilon}_{ij}(\bx_i(s,t),\bx_j(s',t)) \cdot \bff_j(s') \ \mathrm{d} s',
\end{equation}
$\mathsf{G}^\epsilon_{ij}$ is the three-dimensional regularized Blake tensor \cite{blake1971note,cortez2015general} for flow above a no-slip wall. The force per unit length $\bff_i$ can be written as
\begin{equation}
    \bff_i(s,t) = \bff^e_i(s,t) + \sum_{j = i-1}^{i+1} \bff^R_{ij}(s,t).
\end{equation}
It has two contributions: $\bff_i^e$, from the elastic response of the filament backbone, and $\bff_{ij}^R = -\bff_{ji}^R$, which accounts for short-range repulsion between two neighboring filaments. We first focus on the elastic forces $\bff^e$. For a shearable, inextensible, planar rod the elastic force density is given by
\begin{equation}
\bff^e = \partial_s \left(\sigma \uvc{t} + N \uvc{n} \right),
\end{equation}
where $\sigma$ is the force in the tangential direction and $N$ is the force in the normal direction. $\uvc{t},\uvc{n}$ are, respectively, unit tangent and normal vectors to the filament centerline. Since we restrict ourselves to planar deformations of this filament, it is convenient to work in a tangent-angle formulation. We describe the filament by the angle $\phi(s,t)$ made by the centerline with the positive $x$-axis. The unit tangent and normal vectors are then given by $\uvc{t} = \cos \phi \uvc{x} + \sin \phi \uvc{y}$, and $\uvc{n} = -\sin \phi \uvc{x} + \cos \phi \uvc{y}$. 
We model the filament as an Euler elastica in its passive response to bending deformation. It also experiences active bending moments generated by the sliding of microtubules caused by axonemal dynein. The net out-of-plane bending moment is given by
\begin{equation}
\mathbf{M}(s,t) = B \bx_s \times \bx_{ss} - \uvc{z} a \int_s^L \| \bff^{\pm}_{m}(s',t)\| \ \mathrm{d} s',
\end{equation}  
where $B$ is the bending rigidity, $a$ is the diameter of the axoneme, and  $\bff^{\pm}_m$ is the active sliding force per unit length that involves contributions from dynein motors and passive nexin crosslinkers modeled as linear springs.  The sliding force is given as
\begin{equation}
\bff^{\pm}_{m}(s, t)= \pm \bx_s \left[\rho\left(n_{+} F_{+}+n_{-} F_{-}\right)-K \Delta(s,t) \right].
\end{equation}
where $\rho$ is the mean motor density along the filament, $n_{\pm}$ are the two antagonistically operating bound motor populations,  $F_{\pm}$ are the associated loads carried by them, $K$ is the stiffness of the nexin links and $\Delta(s,t) = a (\phi-\phi(s=0))$ is the relative sliding displacement between two microtubule doublets. Moment balance in the out-of-plane direction results in a single scalar equation
\begin{equation}
N = -B \phi_{ss} - a f_m.
\end{equation}
In the above equation we have used the fact that for a planar filament $\bx_s \times \bx_{ss} = \phi_s \uvc{z}$. The elastic force density in the problem is then given by
\begin{equation}
\bff^e = \partial_s \left[\sigma \uvc{t} - \left(B \phi_{ss} + a f_m \right) \uvc{n} \right].
\end{equation}
The tension $\sigma(s,t)$ acts as a Lagrange multiplier to enforce the contraint of inextensibility. We solve for the tension using the fact $\partial_t (\bx_s \cdot \bx_s) = 0$ \cite{tornberg2004simulating,chakrabarti2019spontaneous}.

The bound motor population evolves according to the first-order kinetics $\partial_t n_{\pm} = \pi_{\pm} - \epsilon_{\pm}$, where $\pi_{\pm}$ and $\epsilon_{\pm}$ are attachment and detachment rates, respectively, of the relevant group of motors. The attachment rate is proportional to the fraction of unbound motors and the detachment rate depends exponentially on the carried load. We use a linear force-velocity relationship for the carried load $F_{\pm} =\pm f_0 (1\mp\Delta_t/v_0)$, where $f_0$ is the stall force of the dynein motor and $v_0$ is the characteristic velocity at which the load-carrying capacity is reduced to zero. With this, the attachment and detachment rates can be written as
\begin{align}
\pi_{\pm} &= \pi_0 (1-n_{\pm}), \\
\epsilon_{\pm} &= \epsilon_0 n_{\pm} P_{\pm} \exp \left[ \frac{|F_{\pm}|}{f_c} \right],
\end{align}
where $\pi_0$ and $\epsilon_0$ are characteristic time scales for attachment and detachment, respectively, and $f_c$ is a characteristic force scale above which the motors detach exponentially fast. We also introduce two phenomenological coefficients $P_{\pm}$ that are periodic functions of dynamical variables of the problem. These coefficients are necessary to break the structural symmetry of the axoneme and generate asymmetric beating patterns of cilia. For all the simulations presented in the paper with asymmetric beating pattern, we choose: $P_+ = \exp\left(\tilde{a} \dot{\Delta} \sin \phi_s/v_0\right)$ and $P_{-} = \exp\left( \tilde{b} \dot{\Delta} \sin \Delta /v_0\right)$, where $\tilde{a} = 0.8 \bar{f}$ and $\tilde{b} = -0.6 \bar{f}$ with $\bar{f} = f_0/f_c$.  For symmetric oscillations we set $P_{\pm} = 1$. We emphasize that the precise choice of the phenomenological coefficients does not change any of our findings as long as they break the structural symmetry of the axoneme (see SM). 
Scaling lengths by $L$, sliding displacement
 by $a$, time by the correlation timescale $\tau_0 = 1/(\pi_0 + \epsilon_0)$,
elastic forces by $B/L^2$
, and motor loads by $\rho f_0$ reveals four
dimensionless groups, of which two are of primary interest:
 (i) the sperm number $\mathrm{Sp}=L\left(8 \pi \nu / B \tau_{0}\right)^{1 / 4}$, which compares the relaxation time of a bending mode to
 the motor correlation time; and (ii) the activity number $\mu_a = a \rho f_0 L^2/B$, which compares motor-induced sliding forces to characteristic elastic forces. The two other dimensionless groups
are $\mu=K a^{2} L^{2} / B$   and  $\zeta=a /\left(v_{0} \tau_{0}\right)$ \cite{chakrabarti2019spontaneous}. The dimensionless parameter values used for the simulations are provided in the supporting information. With these scalings, the dimensionless force density is given by
\begin{equation}
    \bff^e = \left[\partial_s \sigma + \phi_s \left(\phi_{ss}+\mu_a f_m \right) \right] \uvc{t} + \left[\sigma \phi_s  - \left(\phi_{sss} + \mu_a \partial_s f_m \right) \right] \uvc{n}.
\end{equation}
The dimensionless evolution equation for the filament centerline follows
\begin{equation}
    \mathrm{Sp}^4 \ \partial_t \bx_i(s,t) - \bu^d(\bx_i(s,t)) =  \mathcal{M} \cdot \bff_i(s,t).
\end{equation}
Finally, the short-range repulsion force between two neighboring filaments in contact has the following form:
\begin{equation}
\bff_{ij}^R(s) = A \left(\frac{\epsilon_d}{d}\right)^{12} \uvc{n}_{ij} \delta_{\varepsilon}(s),
\end{equation} 
where $A = 0.05$, $\epsilon_d = 3 \Delta s$, $d$ is the distance between the two points in contact, $\uvc{n}_{ij}$ is the unit vector joining the two points, and $\delta_{\varepsilon}$ is a regularized delta function that spreads the force over four neighboring nodes. The pair repulsion is only activated  when $d \le \epsilon_d$. 

We discretize the governing equations using a second-order accurate finite difference scheme and solve it using an implicit-explicit second-order accurate time marching scheme that follows \cite{chakrabarti2019spontaneous}. For all the simulations presented here we use $n = 64$ discretization points for the arc length, which means $\Delta s \approx 0.0153$. The time step is set to $\Delta t = 8 \times 10^{-4}$ and is adaptively changed, checking for close contacts of filaments. The regularization parameter for the interaction kernel is set to $\epsilon = 0.005-0.01$. To facilitate fast computations, the interactions between filaments in a periodic box are computed using the fast multipole method \cite{yan2018flexibly,yan2020kernel}.  

\vspace*{1mm}

\noindent \textsf{\textbf{Rower model and simulations:}}  We use a third-order Runge-Kutta method to integrate \eqref{eq:rowev}. To ensure the rowers stay within their tracks, we use a stiff harmonic potential at the switching points. This ensures minimal deviations of the rowers.

\vspace*{1mm}

\noindent \textsf{\textbf{Phase definition:}} To compute the phase of an active filament we first define a continuous, periodic time series $\beta(t) = x(s=1/2,t)$, where $x$ is the horizontal position of the filament. We obtain an analytic continuation of the series as $\zeta(t) = \beta(t) + \mathrm{i} \hat{\beta}(t)$, where 
\begin{equation}
\hat{\beta}(t) = \frac{1}{\pi} \mathrm{p.v} \int_{-\infty}^{\infty} \frac{\beta{\tau}}{t-\tau} \ \mathrm{d} \tau
\end{equation}
is the Hilbert transform of the time series. We then define the filament phase as $\psi(t) = \arctan(\hat{\beta}/\beta)$. For the rowers, we define the phase $\theta_i$ following \cite{wollin2011metachronal} as
\begin{equation}
\theta_{i}=2 \pi n_{i}+\frac{\pi}{2} \sigma_{i} x_{i}+\left\{\begin{array}{ll}
0 & \text { if } x_{i} \in[0,0.5) \wedge \sigma_{i}=1 \\
\pi & \text { if } x_{i} \in[0.5,-0.5) \wedge \sigma_{i}=-1 \\
2 \pi & \text { if } x_{i} \in[-0.5,0) \wedge \sigma_{i}=1.
\end{array}\right.
\end{equation}
We increase $n_i \in \mathbb{Z}$ by unity after each complete oscillation.

\section*{Acknowledgements}

The authors thank David Stein for illuminating discussions and helpful feedback, as well as Robert Blackwell and Wen Yan for help with numerical simulations. MJS acknowledges support by the National Science Foundation under awards DMR- 1420073 (NYU MRSEC) and DMR-2004469.


\section*{Competing interests}
The authors declare no competing interests.

\bibliography{bibcilia}

\end{document}